\documentclass[aps,prl,twocolumn,groupedaddress]{revtex4}

\usepackage{hyperref}
\usepackage{graphicx}
\usepackage{bm}

\begin{document}


\def\hslash{\hbar}
\def\imag{i}
\def\grad{\vec{\nabla}}
\def\div{\vec{\nabla}\cdot}
\def\curl{\vec{\nabla}\times}
\def\DDt{\frac{d}{dt}}
\def\ddt{\frac{\partial}{\partial t}}
\def\ddx{\frac{\partial}{\partial x}}
\def\ddy{\frac{\partial}{\partial y}}
\def\lap{\nabla^{2}}
\def\divv{\vec{\nabla}\cdot\vec{v}}
\def\gradS{\vec{\nabla}S}
\def\vvec{\vec{v}}
\def\wc{\omega_{c}}
\def\<{\langle}
\def\>{\rangle}
\def\Tr{{\rm Tr}}
\def\Csch{{\rm csch}}
\def\Coth{{\rm coth}}
\def\Tanh{{\rm tanh}}
\def\g2{g^{(2)}}
\newcommand{\al}{\alpha}
\newcommand{\la}{\lambda}
\newcommand{\del}{\delta}
\newcommand{\om}{\omega}
\newcommand{\ep}{\epsilon}
\newcommand{\pd}{\partial}
\newcommand{\bra}{\langle}
\newcommand{\ket}{\rangle}
\newcommand{\bbra}{\langle \langle}
\newcommand{\kket}{\rangle \rangle}
\newcommand{\non}{\nonumber}
\newcommand{\be}{\begin{equation}}
\newcommand{\ee}{\end{equation}}
\newcommand{\bea}{\begin{eqnarray}}
\newcommand{\eea}{\end{eqnarray}}

\title{Hamiltonian approach for the wave packet dynamics: Beyond Gaussian wave functions}

\author{Andrey Pereverzev}
\email{aperever@mail.uh.edu}
\affiliation{Department of Chemistry and Texas Center for Superconductivity, 
University of Houston \\ Houston, TX 77204}

\author{Eric R. Bittner}
\thanks{J. S. Guggenheim Fellow, 2007-2008}
\email[email:]{bittner@un.edu}
\affiliation{Department of Chemistry and Texas Center for Superconductivity, 
University of Houston \\ Houston, TX 77204}

\date{\today}

\begin{abstract}
It is well known that the Gaussian wave packet dynamics can be written  in terms of Hamilton equations in the extended phase space that is twice as large as in the corresponding classical 
system. We construct  several generalizations of this approach that include non-Gausssian wave packets. These generalizations lead to the further extension of the phase space while retaining the Hamilton structure of the equations of motion. We compare the Gaussian dynamics with these non-Gaussian extensions for a particle with the quartic potential. 
\end{abstract}


\maketitle
Among various semi-classical approaches the Gaussian wave packet dynamics has been particularly successful and extensively used by  physicists and chemists \cite{Child, Heller1,Cooper,Kramer,Kohen,Ando}.
An especially attractive Hamiltonian formulation of this method was developed by a number of authors \cite{Jackiw, Patt,Cooper,Tsue1,Tsue2}. In this formulation the Gaussian wave packet dynamics is recast into the classical-like dynamics that obeys Hamilton equations in the extended phase space that is twice as large as in the corresponding classical problem. The Hamiltonian approach offers a number of advantages over the non-Hamiltonian formulations \cite{Patt}. Possibly the most important among them is that the existence of the classical-like Hamiltonian insures the stability of trajectories in the generalized phase space and well-defined quantum dynamics. A number of extensions of the Gaussian wave packet approach that include non-Gaussian wave functions either explicitly or implicitly were developed \cite{Heller2,Prezhdo,Pere}. Generally, however, these extensions do not preserve the Hamiltonian structure of the equations of motion and, therefore, do not possess its advantages. The purpose of this paper is to develop a natural extension of the Gaussian wave packet dynamics that preserves the Hamiltonian form for the equations of motion.

The Hamiltonian formulation for both Gaussian wave packet dynamics as well as its non-Gaussian extensions can the most easily be obtained through the time dependent variational principle. In this approach one introduces the functional  
\be
\Gamma=
\int dt(i \hbar \bra \psi (t)| \frac{\pd}{\pd t}|\psi(t)\ket-\bra \psi (t)|H|\psi (t) \ket). \label{Functional}
\ee
The requirement that $\delta\Gamma=0$ against independent variations of $\bra \psi (t)|$ and $|\psi(t)\ket$ leads to the Schr{\"o}dinger equation and its complex conjugate. 
Various approximation schemes can then be constructed by restricting $|\psi\ket$'s to certain classes of functions that depend on a finite number of time dependent parameters. In order to obtain Hamilton equations the parameters should be chosen in such a way that they satisfy the so-called canonicity conditions \cite{Marumori}. An alternative but related approach that does not introduces the canonicity conditions and leads to generalized Hamilton equations is developed in Ref. \cite{Kramer}. The derivation of the Hamilton equations for the Gaussian wave packets become very transparent if,  rather then writing  them in a specific representation, one uses their generic form as squeezed coherent states \cite{Tsue1,Tsue2}. Let us show how Hamilton equations are obtained for the Gaussian wave packets. These results will help us to demonstrate easily how to  obtain the non-Gaussian generalizations.

The Gaussian state is written 
 in terms of the displacement  and squeezing  operators  $D(\alpha)$ and $S(\beta)$ as
\be
|G(\alpha,\beta)\ket=D(\alpha)S(\beta)|0\ket, \label{Gau}
\ee
where $|0\ket$ is the ground state of the harmonic oscillator and displacement and squeezing operators are written in terms of creation and annihilation operators $a$ and 
$a^{\dagger}$ as
\be
D(\alpha)=\exp(\alpha a^{\dagger}-\alpha^*a),\quad
S(\beta)=\exp\left(\frac{\beta}{2} {a^{\dagger}}^2-\frac{\beta^*}{2}  a^2\right).
\ee
The state $|G(\alpha,\beta)\ket$ depends on four real time-dependent parameters. These trial states are inserted in the functional (\ref{Functional}) which will now be expressed 
in terms of these four parameters and their time derivatives. We then need to choose new parameters that satisfy the canonicity conditions. The choice of these new canonical parameters is completely determined by the differential part of the functional (\ref{Functional}).
Writing parameter $\alpha$ as $r_1\exp(i\varphi_1)$ and parameter $\beta$ as $r_2\exp(i\varphi_2)$ and using the properties of $a$ and $a^{\dagger}$ one can verify the following relations


\bea
\frac{\pd }{\pd r_{1}}
DS&=&DS\Big(a^{\dagger}(e^{i\varphi_1}\cosh r_2 -
e^{-i(\varphi_1-\varphi_2)}\sinh r_2) \non \\
& &-a\,( e^{-i\varphi_1}\cosh r_2-
e^{i(\varphi_1-\varphi_2)}\sinh r_2) \Big), \label{Eq1}\\
\frac{\pd }{\pd \varphi_{1}}
DS&=&iDS\Big(r_1^2+a^{\dagger}r_1( e^{i\varphi_1}\cosh r_2+
e^{-i(\varphi_1-\varphi_2)}\sinh r_2)\non\\ 
& &+a\, r_1(e^{-i\varphi_1}\cosh r_2+
 e^{i(\varphi_1-\varphi_2)}\sinh r_2)\Big),\label{Eq2}\\
\frac{\pd }{\pd r_{2}}
DS&=&\frac{1}{2}DS
\Big({a^{\dagger}}^{2}e^{i\varphi_{2}}-{a}^{2}e^{-i\varphi_{2}}\Big),\label{Eq3}  \\
\frac{\pd }{\pd \varphi_2}
DS&=&\frac{i}{2}DS\Big(({a^{\dagger}}^{2}e^{i\varphi_{2}}
+a^2e^{-i\varphi_{2}})\sinh r_{2}\cosh r_{2} \non \\  
& &+(2a^{\dagger}a+1)\sinh^{2} r_2\Big). \label{Eq4}
\eea
In the case of the Gaussian trial state (\ref{Gau}) when calculating the time derivatives for the approximating functional we need to multiply these expressions from the left by $(DS)^{-1}$ and then average them over the vacuum state. Performing this averaging we obtain
\bea
\bra G|\frac{\pd }{\pd r_{1}}|G\ket&=&\bra G|\frac{\pd }{\pd r_{2}}|G\ket=0 \label{zeroder}\\ 
\bra G|\frac{\pd }{\pd \varphi_{1}}|G\ket&=&i r_1^2, \quad
\bra G|\frac{\pd }{\pd \varphi_{2}}|G\ket=\frac{i}{2}\sinh^2\!r_2.
\eea
Using these results and 
introducing the new variables $J_1=r_1^2$ and $J_2=\frac{1}{2}\sinh^{2}\! r_2$ while keeping the original $\varphi_{i}$'s we obtain  the following form of the functional (\ref{Functional}) for the Gaussian state (\ref{Gau})
\be
\Gamma=\int dt\big(-\sum_{i=1}^{2}J_i\dot\varphi_i-{\cal H}( \{J_i\}, \{\varphi_i\})\big),\label{Functional2}
\ee
where ${\cal H}( \{J_i\}, \{\varphi_i\})=\bra G|H|G\ket$ and $\hbar$ is set to unity. Parameters $J_{i}$ and $\varphi_{i}$ satisfy the canonicity conditions. The requirement that $\delta\Gamma=0$ leads to the following Hamilton equations
\be
\dot\varphi_i=-\frac{\pd {\cal H}}{\pd J_i}, 
\qquad \dot J_i=\frac{\pd {\cal H}}{\pd \varphi_i}.
\ee

We now want to generalize these results to functions that are not necessarily Gaussian.
An obvious generalization of the squeezed state of Eq. (\ref{Gau}) is the state of the form
$|F\ket=D(\alpha)S(\beta)|\xi\ket$ where $|\xi\ket$ is a state depending on an even number of real parameters.
Note that when we consider the variation of functional $\Gamma$ with  function $|F\ket$ instead of 
$|G\ket$ we can still use Eqs. (\ref{Eq1}-\ref{Eq4}) for the derivatives with respect to 
$r_1, \varphi_1, r_2$ and $\varphi_2$.
To ensure that we end up with the Hamilton equations, we also want to preserve the property that the derivatives with respect to $r_1$ and $r_2$ that appear in Eqs. (\ref{Eq1},\ref{Eq3}) vanish after averaging as in Eq. (\ref{zeroder}) . This implies that the state $|\xi\ket$ must be such that the averages of $a$, $a^{\dagger}$, $a^2$, and ${a^{\dagger}}^2$ are equal to zero. This is valid for any state that is a superposition of the number states with occupation numbers that differ at least by $3$. In this paper we consider
 trial states $|\xi\ket$  of the following general form 
\be
|\xi\ket
=\sum_{m=0}^{M} c_m |3m\ket, \label{3state}
\ee
that are assumed to be normalized with $M$ being an integer that can range from $1$ to infinity. 
Thus, instead of the squeezed and displaced vacuum as in the case of the Gaussian state, we are now dealing with some squeezed and displaced superposition of states with occupation numbers that are multiples of $3$.
Recall that the squeezing operator acting on the vacuum produces states that are superpositions of number  states  with occupation numbers that are multiples of $2$. 
In this sense the state $|F\ket$ appears as a natural generalization 
of $|G\ket$.  We now have to choose some  parametrization
of coefficients $c_m$ that satisfies the canonicity conditions. Here we consider two possible types of parametrization for the function $|\xi\ket$.

A simple two parameter parametrization of the state (\ref{3state}) that is close in spirit to the original Gaussian state is obtained by writing the state $|\xi\ket$ as
\be
|\xi(\gamma)\ket
=\frac{1}{\sqrt{N(\gamma)}}\sum_{m=0}^{\infty} b_m \gamma^{m} |3m\ket, \label{gammastate}
\ee
where coefficients $b_m$ are fixed and the complex parameter $\gamma$ is allowed to change. The normalization constant $N$ is given by
$
N(\gamma)=\sum_{0}^{\infty}|b_m|^2|\gamma|^{2m}.
$
Eq. (\ref{gammastate}) is the general form for several classes of generalized squeezed and coherent states found in the literature, such as various multiphoton squeezed states  \cite{D'Ari} or eigenstates of operator $a^3$ \cite{Dodo,Buz,Nieto}. All these states are distinguished  only by a particular choice of coefficients $b_m$ in Eq. (\ref{gammastate}). Since  these states are always normalized and reduce to the vacuum state for $\gamma=0$ they can be written in terms of some two-parameter unitary operator $T(\gamma)$ acting on the vacuum. The total trial state then has the form  $|F(\alpha,\beta)\ket=D(\alpha)S(\beta)T(\gamma)|0\ket$ that looks like a direct generalization of the squeezed coherent state $D(\alpha)S(\beta)|0\ket$. The explicit form of $T(\gamma)$ is only known for some special choices of $b_m$'s \cite{D'Ari} and for this reason we will use the form (\ref{gammastate}) instead of $T(\gamma)|0\ket$ in the further analysis. Writing $\gamma$ as $r_3\exp{i\varphi_3}$ one easily verifies that 
\bea
\bra F|\frac{\pd }{\pd r_{3}}|F\ket&=&0, \label{Eq5}\\
\bra F|\frac{\pd }{\pd \varphi_{3}}
|F\ket&=&\frac{i}{N(\gamma)}
\sum_{m=0}^{\infty}m |c_m|^2|\gamma |^{2m}\non \\
&=&\frac{i}{3}\bra F|(a^{\dagger}a) |F\ket \equiv iK(r_3). \label{Eq6}
\eea
Using these relations as well as Eqs.(\ref{Eq1}-\ref{Eq4}) we obtain the new variables that satisfy the canonicity conditions and Hamilton equations
\be
J_1=r_1^2,\qquad J_2=\frac{1}{2}(2K(r_3)+1)\sinh^{2}\! r_2,\qquad J_3=K(r_3),\label{NewJ}
\ee 
with the original $\varphi_i$'s. The ``classical" Hamiltonian ${\cal H}( \{J_i\}, \{\varphi_i\})$ is given by  $\bra F|H|F\ket$.
Application of these results to a specific Hamiltonian
will require a particular choice for the form of coefficients $b_m$ in Eq. (\ref{gammastate}). Given a Hamiltonian, this choice can be guided by both physical plausibility and mathematical manageability. In particular, we should be  able to rewrite ${\cal H}$ in terms of $J_i$'s 
rather then the original $r_i$'s by using Eqs. (\ref{NewJ}). Generally, this  implies that we should be able to solve the last of Eqs. (\ref{NewJ}) for $r_3$. In the example below we take $c_m$'s to be: $c_m=1/\sqrt{m!}$. The relation of the corresponding state to the algebraic properties of the generalized three photon creation and annihilation operators is discussed in Ref. \cite{D'Ari}. We make this particular choice here primarily because of its mathematical convenience, since in this case $K(r_3)={r_3}^2$ and, therefore, $r_3$ is easily expressed in terms of $J_3$. To distinguish this state from the other trial states we denote it $|F_1\ket$.

Another, more detailed parametrization is achieved by treating coefficients $c_m$ 
in Eq. (\ref{3state})  themselves as parameters. To this end we write each of $c_m$'s (for $m>0$) as $R_m\exp(i\phi_m)$. By the normalization requirement $c_0$ is then given by $c_0=\sqrt{1-\sum_{1}^{M}|c_m|^2}$. It is easy to verify that
\be
\bra F|\frac{\pd }{\pd R_{i}}|F\ket=0, \qquad
\bra F|\frac{\pd }{\pd \phi_{i}}|F\ket={i}R_i^2
\ee
Using these relations as well as Eqs.(\ref{Eq1}-\ref{Eq4}) we obtain 
the new variables $J_i$ that are expressed through the original parameters as follows
\be
J_1=r_1^2,\qquad J_2=\frac{1}{2}(1+2\sum_{m=1}^{\infty}3m|R_m|^2)\sinh^{2}\! r_2,
\ee
and for  $i\geq 3$
\be
J_i=R_{i-2}^2, \qquad \varphi_i=\phi_{i-2}. 
\ee
The original $\varphi_{1}$ and $\varphi_{2}$ are the conjugate variables of $J_1$ and $J_2$. The value of integer $M$ determines by how much the phase space is extended compared to the Gaussian case. In the example below we consider the simple cases of $M=1$ (six dimensional phase space) and $M=2$ (eight dimensional phase space). We will denote the corresponding trial states
by $|F_2\ket$ and $|F_3\ket$.  An interesting open question that we do not consider here  is how well can an arbitrary function be approximated by this type of the trial state when $M$ is allowed to be infinite.

Let us now compare these non-Gaussian extensions with the Gaussian wave packet dynamics
in the case of the model Hamiltonian
\be
H=\frac{p^2}{2}+\frac{a x^2}{2}+\frac{\lambda x^4}{4}. \label{quartic}
\ee
We will take the value of  $\lambda=1$ and, therefore, the Hamiltonian is bounded from below. We will consider two choices for $a$: $a=1$ and  $a=-1$. The latter case corresponds to a two-well potential with the two minima located at $\pm 1$. 
In order to make the comparison with the Gaussian approach we have to choose the initial state as a Gaussian one given by Eq. (\ref{Gau}). We consider the same initial state for both choices of $a$ with parameters $\alpha=1/\sqrt{2}$ and $\beta=0.1$. This correspond to a slightly squeezed state with the average coordinate equal to one and zero average momentum.
With these parameters the initial expectation values of the quadratic and quartic potential terms in Eq. (\ref{quartic}) are of the same order.

\begin{figure}
 \includegraphics[width=\columnwidth]{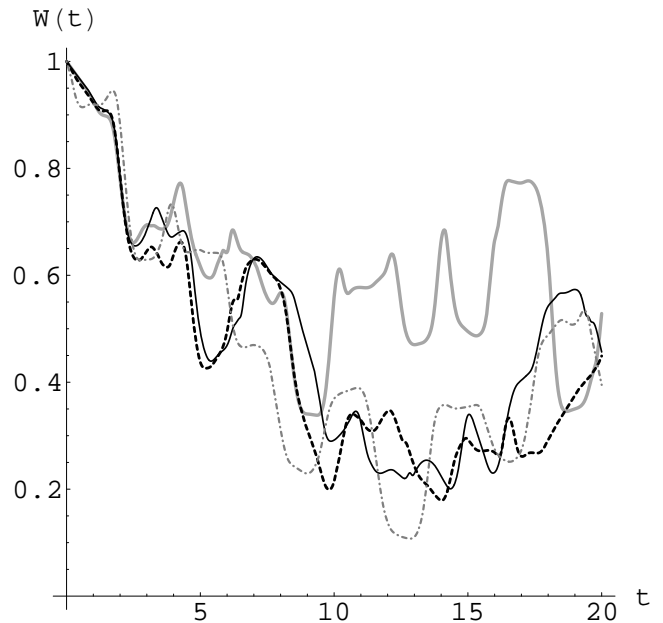} 
 \caption{\label{Overlap2} $W(t)$ for $|G\ket$ (gray dots and dashes),  $|F_1\ket$  (solid gray), $|F_2\ket$ (black dashes), and $|F_3\ket$ (solid black) when $a=-1$ .}
 \end{figure}

\begin{figure}
 \includegraphics[width=\columnwidth]{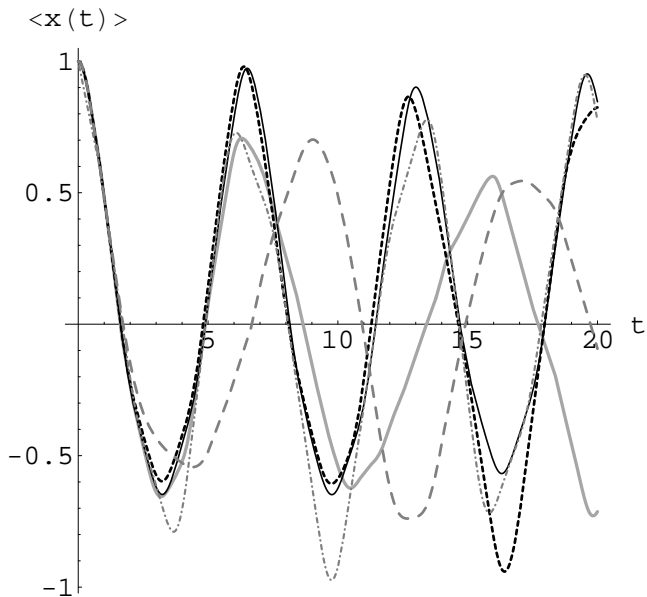} 
 \caption{\label{avex} $\bra x(t)\ket$ for the two-well potential ($a=-1$) for the exact wave function (long gray dashes), $|G\ket$ (gray dots and dashes),  $|F_1\ket$  (solid gray), $|F_2\ket$ (short black dashes), and $|F_3\ket$ (solid black).}
 \end{figure}

First we consider the case of $a=-1$. The main quantity of interest for us is $W(t)$, the squared overlap between the exact wave function and its approximations given by  $|G\ket, |F_1\ket, |F_2\ket$, and $|F_3\ket$ (Fig. \ref{Overlap2}). The exact wave function is calculated by the numerical solution of the Schr{\"o}dinger equation. To give the reader a better feeling of the time scales, Fig. \ref{avex} shows the time evolution of the average coordinate $\bra x(t)\ket$ over the same time interval.
We can see from Fig. \ref{Overlap2} that for short times (up to about $1$ time unit) all three non-Gaussian extensions give better overlap with the exact wave function then the Gaussian approximation. For longer times, however, the behavior of  $W(t)$ for different approximations becomes more complicated. Depending on a specific time {\it each} of the approximating wave functions (with the exception of $|F_2\ket$) can give the best overlap with the exact wave function.
When averaged over the time interval of the Figure (20 time units) the 
values of the squared overlap are the following: $0.466$ for $|G\ket$, $0.612$ for $|F_1\ket$, $0.450$ for $|F_2\ket$, and $0.492$ for $|F_3\ket$. Thus, on average for this time interval $|F_1\ket$ provides a much better description then the Gaussian wave function, $|F_3\ket$ performs slightly better, and 
$|F_2\ket$ gives {\it worse} results then the Gaussian. These results are in agreement with 
Fig. (\ref{avex}) which shows that $|F_1\ket$ generally reproduces exact $\bra x(t)\ket$ better then the other functions.

\begin{figure}
 \includegraphics[width=\columnwidth]{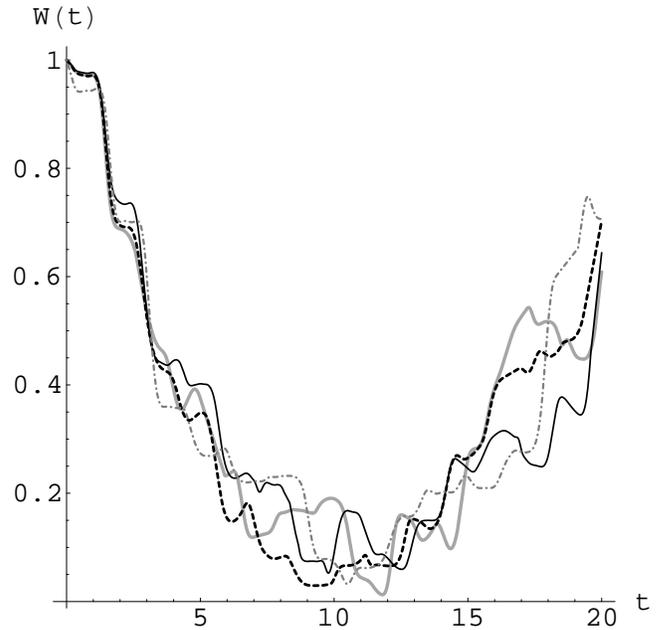} 
 \caption{\label{Overlap1} $W(t)$ for $|G\ket$ (gray dots and dashes),  $|F_1\ket$  (solid gray), $|F_2\ket$ (black dashes), and $|F_3\ket$ (solid black) when $a=1$.}
\end{figure}

The squared overlaps for the case of $a=1$ is shown on Figure \ref{Overlap1}.
We see again that for up to about $1$ time unit the non-Gaussian wave functions are closer to the exact one then the Gaussian wave packet. For longer times no trial function seems to offer a definite advantage.  $W(t)$ averaged over $20$ time units has the following  values: $0.353$ for $|G\ket$, $0.353$ for $|F_1\ket$, $0.336$ for $|F_2\ket$, and $0.338$ for $|F_3\ket$. For this case both
 $|F_2\ket$ and  $|F_3\ket$ give slightly worse averaged $W(t)$ then the Gaussian case.
 
To sum up, in both cases ($a=\pm1$) the short time dynamics is improved by using extended wave functions. For longer times  $|F_1\ket$ performs at least as good as  $|G\ket$ or better, while 
$|F_2\ket$ and $|F_3\ket$ show no definite improvement over  $|G\ket$ and can even perform worse then $|G\ket$. This long time behavior seems surprising. Indeed, one usually expects that variational approaches give results that become progressively better when the number of variational parameters increases. We do not offer an explanation of this kind of behavior in this paper. It is possible that the reason for this should be sought in the interpretation of the variational principle when applied to the functionals that linearly depend on the time derivative of the field. When performing the variation of the functional (\ref{Functional}) to derive the Schr{\"o}dinger equation we assume that we can independently vary $|\psi\ket$ at both time limits. In the resulting Schr{\"o}dinger equation, however, the initial wave function uniquely  determines the final one.

To conclude, we constructed several generalizations of the Gaussian wave packet dynamics that include non-Gaussian wave functions and can be formulated in terms of the classical dynamics in the extended phase space. Applied to a model system these approaches give a better description of the exact wave function for short times. Understanding the peculiarities of the longer time behavior as well as further extensions of these approximations require additional research.   

\begin{acknowledgments}
This work was funded in part by grants from the National Science
Foundation (CHE-0712981), the Robert A. Welch Foundation (E-1337), and the Texas Learning and Computation Center (TCL$^2$).
\end{acknowledgments}


\begin{thebibliography}{18}
\expandafter\ifx\csname natexlab\endcsname\relax\def\natexlab#1{#1}\fi
\expandafter\ifx\csname bibnamefont\endcsname\relax
  \def\bibnamefont#1{#1}\fi
\expandafter\ifx\csname bibfnamefont\endcsname\relax
  \def\bibfnamefont#1{#1}\fi
\expandafter\ifx\csname citenamefont\endcsname\relax
  \def\citenamefont#1{#1}\fi
\expandafter\ifx\csname url\endcsname\relax
  \def\url#1{\texttt{#1}}\fi
\expandafter\ifx\csname urlprefix\endcsname\relax\def\urlprefix{URL }\fi
\providecommand{\bibinfo}[2]{#2}
\providecommand{\eprint}[2][]{\url{#2}}

\bibitem[{\citenamefont{Child}(1991)}]{Child}
\bibinfo{author}{\bibfnamefont{M.~S.} \bibnamefont{Child}},
  \emph{\bibinfo{title}{Semiclassical Mechanics with Molecular Applications}}
  (\bibinfo{publisher}{Clarendon, Oxford}, \bibinfo{year}{1991}).

\bibitem[{\citenamefont{Heller}(1975)}]{Heller1}
\bibinfo{author}{\bibfnamefont{E.~J.} \bibnamefont{Heller}},
  \bibinfo{journal}{J. Chem. Phys.} \textbf{\bibinfo{volume}{62}},
  \bibinfo{pages}{1544} (\bibinfo{year}{1975}).

\bibitem[{\citenamefont{Cooper et~al.}(1987)\citenamefont{Cooper, S.-Y.Pi, and
  Stancioff}}]{Cooper}
\bibinfo{author}{\bibfnamefont{F.}~\bibnamefont{Cooper}},
  \bibinfo{author}{\bibnamefont{S.-Y.Pi}}, \bibnamefont{and}
  \bibinfo{author}{\bibfnamefont{P.~N.} \bibnamefont{Stancioff}},
  \bibinfo{journal}{Phys. Rev. D} \textbf{\bibinfo{volume}{34}},
  \bibinfo{pages}{3831} (\bibinfo{year}{1986}).

\bibitem[{\citenamefont{Kramer and Saraceno}(1981)}]{Kramer}
\bibinfo{author}{\bibfnamefont{P.}~\bibnamefont{Kramer}} \bibnamefont{and}
  \bibinfo{author}{\bibfnamefont{M.}~\bibnamefont{Saraceno}},
  \emph{\bibinfo{title}{Geometry of the Time-Dependent Variational Principle in
  Quantum Mechanics}} (\bibinfo{publisher}{Springer-Verlag, Berlin},
  \bibinfo{year}{1981}).

\bibitem[{\citenamefont{Kohen and Tannor}(1997)}]{Kohen}
\bibinfo{author}{\bibfnamefont{D.}~\bibnamefont{Kohen}} \bibnamefont{and}
  \bibinfo{author}{\bibfnamefont{D.~J.} \bibnamefont{Tannor}},
  \bibinfo{journal}{J. Chem. Phys.} \textbf{\bibinfo{volume}{107}},
  \bibinfo{pages}{5141} (\bibinfo{year}{1997}).

\bibitem[{\citenamefont{Ando}(2004)}]{Ando}
\bibinfo{author}{\bibfnamefont{K.}~\bibnamefont{Ando}}, \bibinfo{journal}{J.
  Chem. Phys.} \textbf{\bibinfo{volume}{121}}, \bibinfo{pages}{7136}
  (\bibinfo{year}{2004}).

\bibitem[{\citenamefont{Jackiw and Kerman}(1979)}]{Jackiw}
\bibinfo{author}{\bibfnamefont{R.}~\bibnamefont{Jackiw}} \bibnamefont{and}
  \bibinfo{author}{\bibfnamefont{A.}~\bibnamefont{Kerman}},
  \bibinfo{journal}{Phys. Lett. A} \textbf{\bibinfo{volume}{71}},
  \bibinfo{pages}{158} (\bibinfo{year}{1979}).

\bibitem[{\citenamefont{Pattanayak and Schieve}(1994)}]{Patt}
\bibinfo{author}{\bibfnamefont{A.~K.} \bibnamefont{Pattanayak}}
  \bibnamefont{and} \bibinfo{author}{\bibfnamefont{W.~C.}
  \bibnamefont{Schieve}}, \bibinfo{journal}{Phys. Rev. E}
  \textbf{\bibinfo{volume}{50}}, \bibinfo{pages}{3601} (\bibinfo{year}{1994}).

\bibitem[{\citenamefont{Tsue and Fujiwara}(1991)}]{Tsue1}
\bibinfo{author}{\bibfnamefont{Y.}~\bibnamefont{Tsue}} \bibnamefont{and}
  \bibinfo{author}{\bibfnamefont{Y.}~\bibnamefont{Fujiwara}},
  \bibinfo{journal}{Prog. Theor. Phys.} \textbf{\bibinfo{volume}{86}},
  \bibinfo{pages}{443} (\bibinfo{year}{1991}).

\bibitem[{\citenamefont{Tsue}(1992)}]{Tsue2}
\bibinfo{author}{\bibfnamefont{Y.}~\bibnamefont{Tsue}}, \bibinfo{journal}{Prog.
  Theor. Phys.} \textbf{\bibinfo{volume}{88}}, \bibinfo{pages}{911}
  (\bibinfo{year}{1992}).

\bibitem[{\citenamefont{Heller}(1976)}]{Heller2}
\bibinfo{author}{\bibfnamefont{E.~J.} \bibnamefont{Heller}},
  \bibinfo{journal}{J. Chem. Phys.} \textbf{\bibinfo{volume}{64}},
  \bibinfo{pages}{63} (\bibinfo{year}{1976}).

\bibitem[{\citenamefont{Prezhdo and Pereverzev}(2000)}]{Prezhdo}
\bibinfo{author}{\bibfnamefont{O.~V.} \bibnamefont{Prezhdo}} \bibnamefont{and}
  \bibinfo{author}{\bibfnamefont{Y.~V.} \bibnamefont{Pereverzev}},
  \bibinfo{journal}{J. Chem Phys.} \textbf{\bibinfo{volume}{113}},
  \bibinfo{pages}{6557} (\bibinfo{year}{2000}).

\bibitem[{\citenamefont{Pereverzev et~al.}(2008)\citenamefont{Pereverzev,
  Pereverzev, and Prezhdo}}]{Pere}
\bibinfo{author}{\bibfnamefont{A.}~\bibnamefont{Pereverzev}},
  \bibinfo{author}{\bibfnamefont{Y.~V.} \bibnamefont{Pereverzev}},
  \bibnamefont{and} \bibinfo{author}{\bibfnamefont{O.~V.}
  \bibnamefont{Prezhdo}}, \bibinfo{journal}{J. Chem. Phys.}
  \textbf{\bibinfo{volume}{128}}, \bibinfo{pages}{134107}
  (\bibinfo{year}{2008}).

\bibitem[{\citenamefont{Marumori et~al.}(1980)\citenamefont{Marumori, Maskawa,
  Sakata, and Kuriyama}}]{Marumori}
\bibinfo{author}{\bibfnamefont{T.}~\bibnamefont{Marumori}},
  \bibinfo{author}{\bibfnamefont{T.}~\bibnamefont{Maskawa}},
  \bibinfo{author}{\bibfnamefont{F.}~\bibnamefont{Sakata}}, \bibnamefont{and}
  \bibinfo{author}{\bibfnamefont{A.}~\bibnamefont{Kuriyama}},
  \bibinfo{journal}{Prog. Theor. Phys.} \textbf{\bibinfo{volume}{64}},
  \bibinfo{pages}{1294} (\bibinfo{year}{1980}).

\bibitem[{\citenamefont{D'Ariano et~al.}(1987)\citenamefont{D'Ariano, Morosi,
  Rasetti, Katriel, and Solomon}}]{D'Ari}
\bibinfo{author}{\bibfnamefont{G.}~\bibnamefont{D'Ariano}},
  \bibinfo{author}{\bibfnamefont{S.}~\bibnamefont{Morosi}},
  \bibinfo{author}{\bibfnamefont{M.}~\bibnamefont{Rasetti}},
  \bibinfo{author}{\bibfnamefont{J.}~\bibnamefont{Katriel}}, \bibnamefont{and}
  \bibinfo{author}{\bibfnamefont{A.~I.} \bibnamefont{Solomon}},
  \bibinfo{journal}{Phys. Rev. D} \textbf{\bibinfo{volume}{36}},
  \bibinfo{pages}{2399} (\bibinfo{year}{1987}).

\bibitem[{\citenamefont{Dodonov et~al.}(1974)\citenamefont{Dodonov, Malkin, and
  Man'ko}}]{Dodo}
\bibinfo{author}{\bibfnamefont{V.~V.} \bibnamefont{Dodonov}},
  \bibinfo{author}{\bibfnamefont{I.~A.} \bibnamefont{Malkin}},
  \bibnamefont{and} \bibinfo{author}{\bibfnamefont{V.~I.}
  \bibnamefont{Man'ko}}, \bibinfo{journal}{Physica}
  \textbf{\bibinfo{volume}{72}}, \bibinfo{pages}{597} (\bibinfo{year}{1974}).

\bibitem[{\citenamefont{Bu{\v z}ek et~al.}(1990)\citenamefont{Bu{\v z}ek, Jex,
  and Quang}}]{Buz}
\bibinfo{author}{\bibfnamefont{V.}~\bibnamefont{Bu{\v z}ek}},
  \bibinfo{author}{\bibfnamefont{I.}~\bibnamefont{Jex}}, \bibnamefont{and}
  \bibinfo{author}{\bibfnamefont{T.}~\bibnamefont{Quang}}, \bibinfo{journal}{J.
  Mod. Opt} \textbf{\bibinfo{volume}{37}}, \bibinfo{pages}{159}
  (\bibinfo{year}{1990}).

\bibitem[{\citenamefont{Nieto and Truax}(2000)}]{Nieto}
\bibinfo{author}{\bibfnamefont{M.~M.} \bibnamefont{Nieto}} \bibnamefont{and}
  \bibinfo{author}{\bibfnamefont{D.~R.} \bibnamefont{Truax}},
  \bibinfo{journal}{Optics Com.} \textbf{\bibinfo{volume}{179}},
  \bibinfo{pages}{197} (\bibinfo{year}{2000}).

\end{thebibliography}
\end{document}